\documentstyle[preprint,tighten,eqsecnum,epsfig,aps]{revtex}
\input{psfig}

\def\gs{\alpha_sN_c/(\pi (n-1))}
\def\beq{\begin{equation}}   \def\eeq{
\end{equation}}

\begin{document}

\title{Large distance behaviour of light cone operator product
in perturbative and nonperturbative QCD regimes}
\author{ B. Blok$^1$\thanks{E-mail:blok@physics.technion.ac.il}
 and L. Frankfurt$^2$\thanks{E-mail:frankfur@lev.tau.ac.il} }
\address{$^1$Department of Physics, Technion -- Israel Institute of
Technology, Haifa 32000, Israel\\[10pt]
 $^2$School of Physics and
Astronomy, Faculty of Natural Sciences, Tel Aviv University,
Israel.} \maketitle

\thispagestyle{empty}

\begin{abstract}
We evaluate the coordinate space dependence of the matrix elements of
the commutator of the electromagnetic and  gluon currents in the vicinity
of the light-cone but at large distances within the parton model,
DGLAP, the resummation approaches to the small x behaviour of DIS
processes, and 
for the Unitarity Bound.
We find that an increase of the
commutator with relative distance $py$ as $\propto
(py)f(py,y^2=t^2-r^2)$ is the generic property of QCD
at small but fixed space-time interval $y^2=t^2-r^2$ in
perturbative and nonperturbative QCD regimes. We explain that the
factor $py$ follows within the dipole model (QCD factorization
theorem) from the properties of Lorents transformation.  The
increase of $f(r)$ disappeares at central impact parameters if
cross section of DIS may achieve the Unitarity Limit. We argue
that such long range forces are hardly consistent with
thermodynamic equilibrium while a Unitarity Limit may signal
equilibration.

Possible implications of this new long range interaction are briefly
discussed.

 \end{abstract}

 \setcounter{page}{1} \section{Introduction}

Commutators of the local currents in the coordinate space play an
important role in the quantum field theory . In particular,they
help to visualize the relationship between the quantum field
theory and statistical mechanics of equilibrium and nonequilibrium
systems. It has been understood that many properties of deep
inelastic processes follow from the operator product expansion. In
particular, the dependence of the product of local currents:
\begin{eqnarray}
<N\vert j_{\mu}(y)j_{\nu}(o)\vert N> &=& (1/y^2)^2\sum_n
p_{\mu}p_{\nu} (py)^n <N\vert O_n(0)\vert N> +\mbox{NLT
terms}\nonumber\\[10pt]
 &=&p_{\mu}p_{\nu} (F(py,y^2)/(y^2)^2) +\mbox{NLT
terms.}\nonumber\\[10pt]
\label{fr1}
\end{eqnarray}
on the space-time interval $y^2=t^2-r^2$ unambigously follows from
this expansion for the leading term \cite{Frishman1} and from
renormalization group. (For certainty we write formulae for the
product of e.m. or gluon currents and neglect by longitudinal
structure functions).  We will show in this paper that the
dependence of operator product on relative distance-$(py)$ at
fixed $y^2$ as
\begin{equation}
F(py,y^2)\propto (py)f(py,y^2)
\end{equation}
with f increasing with $(py)$ follows  from the basic properties of QCD.

\par The actual behaviour of the structure functions of the nucleon at small
Bjorken
$x=Q^2/2pq$
is still a challenging question now as it was 30 years ago.
(Here $-Q^2$ is mass$ ^2$ of an incoming photon).
So a variety  of the new approaches to small
$x$
phenomena were developed such as: the generalization of the QCD
factorization theorem to the amplitudes of the hard diffractive
processes which justifies the dipole approach \cite{BBFS,CFS}; the
derivation within the BFKL approximation \cite{M} of the dipole
approach in the large $N_c$ limit, the resummation of the pQCD
series within the DGLAP approximation \cite{ABF}, the NLO BFKL
approximation and resummation \cite{Ciafaloni,LipF}, the
MacLerran-Venugopalan model \cite{MV}; the eikonal approximation
where a "potential" is evaluated within the DGLAP or BFKL
approximations \cite{JKLW,Levin,Kovchegov},
Unitarity Bound (Black Body Limit) approach \cite{MacDermot}
\par It is well known that the
theoretical description of the high energy processes is
significantly simplified in the coordinate space even if actual
calculations may appear rather cumbersome. The aim of this paper
is to evaluate the amplitudes of the deep-inelastic (DIS)
processes in coordinate space and to visualize the dominant
physics. The knowledge of the space $-$ time evolution of DIS is
especially important for the theoretical description of the RHIC
program of heavy ion collisions, for the QCD part of the LHC
program and for the hunt for the new particles \cite{FS,Wess,FS1}.
\par  It has been demonstrated that DGLAP approximation describes well
increase of  structure functions of a proton \cite{Zeus} observed
by H1 and ZEUS. The experimental data can be fitted as:
\beq
xG_{}(x,Q^2),\,\,\,\,\,F_{2}(x,Q^2) \propto
x^{-\lambda}\label{i1}\eeq with $\lambda\approx 0.25$
\cite{Zeus}. Basic features of hard diffractive (HT) processes
observed by H1 and ZEUS \cite{Zeus} are well described by the QCD
factorization theorem \cite{CFS}. The success of DGLAP at the
energy range covered by HERA is due to the energy-momentum
conservation law restrictions on the possible number of gluons
radiated in the multi Regge kinematics. In the kinematics covered
by HERA this number is equal to 1-2.  At even smaller
$x$
(this corresponds to the LHC kinematics and larger energies) the multiplicity
of radiated gluons may achieve 5-6 and therefore PQCD approximations become
unstable because of the necessity to account for large
$\log(x_{0}/x)$ terms.
So various resummation procedures were suggested \cite{ABF,Ciafaloni}.

\par
Significant cross section of diffraction in DIS observed by H1 and
ZEUS \cite{Zeus} are hardly consistent with the validity of the
leading twist (LT)  approximation at large $Q^2$ but sufficiently
small x. The theoretical analysis shows that the LT approximation
is probably violated in the kinematics which is not far from that
investigated at HERA \cite{Zeus}. Account of the conservation of
the probability  leads to the Unitarity Bound which is the
generalization to DIS of the Froissart limit familiar from the
hadron-hadron collisions . The Unitarity Bound formula shows that
the structure functions of a nucleon may  increase with the energy
as \cite{MacDermot}
\beq F_2,\,\,\,\,xG\propto \log^3(x_{0}/x).\label{i2}\eeq
Increase of the structure functions with the energy follows from
the increase with the energy of the essential impact parameters
factor  $\log^2(x_{0}/x)$
and from the ultraviolet divergence of the renormalization
constants in QCD
-factor  $\log(x_{0}/x)$.
The conservation of the probability permits more moderate
increase of the structure functions at the central impact parameters as
\beq F_2,\,\,\,\, xG \propto \log(x_{0}/x). \label{i3} \eeq

\par It has been understood already in 60-s that the dependence of
the amplitude of the deep-inelastic scattering on $\nu=2pq$ and
photon virtuality -$Q^2$ gives unique possibility to probe the
space-time behaviour of the DIS processes \cite{GIP}. (On the
contrary the amplitudes of the soft QCD processes are always on
mass shell. So it is impossible to compare with the data the
dependence of the commutator of currents on the space-time
interval $y^2=t^2-r^2$ and the relative distance $py$.) Increase
with the energy of the
coherence length in the target rest frame has been suggested in
ref. \cite{GIP}, based on  the analogy with the QED coherent
phenomena in the high energy electron interaction in the medium
\cite{LP,TM}. The formulae for the coherence length :
$l_c\approx 1/2m_Nx$ follows from the properties of the Fourier
transform in \cite{GIP}  since the amplitude of the DIS decreases
with increase of $Q^2$.
Recently, the calculation of ref. \cite{KS} found that account for the
pQCD radiation leads to
significantly smaller (but still  increasing fast with the energy)
coherent length as compared to that found within the parton model.
The first theoretical analysis of the space-time evolution of high
energy processes in a quantum field theory was given by Gribov
\cite{gribov} and applied for the calculation  of the nuclear
shadowing in the electron -nucleus interactions \cite{GribovShad}.
\par The dependence of the DIS amplitude on the light-cone interval
$y^2$ was studied  extensively in the parton model
\cite{Ioffe,IKL,Frishman1,Frishman2}.
It was also studied in the framework of a Regge ansats for small $x_B$
structure function behaviour \cite{Frishman1,Frishman2} .
\par   The aim of the paper is to show that for for sufficiently small
but fixed $y^2$ the product of the currents  is the increasing
function of $py$ -
distance (time),  within both  the leading log (LL) or resummation
approaches to pQCD $\propto (py)f(py,y^2)$ where f is increasing
function of $py$ at fixed $y^2$. Moreover, we shall see that the
increase of the product of the color neutral currents with $(py)$
is valid for the phenomenological structure functions, describing
HERA data. Increase of f disappears in the Unitarity Bound
approximation but for the fixed impact parameters only. In
particular, in perturbative QCD\beq
 F(py,y^2)\propto \theta(y^2)/(y^2)^2(\alpha_sN_c/\pi)^{1/4}
\displaystyle{\frac{\log(Q_0^2y^2)^{1/4}}{\log(py)^{3/4}}}
\exp(2\sqrt{\alpha_sN_c/\pi \log(py)\log(Q_0^2y^2)})
 \label{end1a}
 \eeq
 and in the
black limit \beq F(py,y^2)\propto \theta (y^2)
\log^3((py))/(y^2)^2) +\mbox{ peripheral terms} \label{53i} \eeq

\par
Note that
similar
increase is characteristic for turbulence. It is well known that
the velocity-velocity correlator increases with distance in such a
system for the case of homogeneous turbulence, i.e. for the scales
much smaller than the scale of the entire system.
\cite{LLG}. In turbulence such a behaviour arises because
 the same piece of matter reveals itself in different points.
Similarly in the deep-inelastic scattering  in the target rest
frame the same dipole reveals itself in different space-time
points as a consequence of the large coherence length. This
explains factor r in the matrix element of commutator. Increase of
$f(r,y^2)$ with r ($r\sim t$) indicates that produced perturbative
system is far from the thermodynamic equilibrium. Some caution is
to the point: our interest is in the distances less or comparable
with coherence length. At larger distances deduced formulae are
hardly applicable at the distances $\gg l_c$ where nonperturbative
phenomena: confinement of color and phenomenon of spontaneously
broken chiral symmetry should be important. Discussion of this
important question is beyond the scope of this publication. Note
however that the regime leading to the Unitarity Bound corresponds
to f not increasing with distance which is a hint for the
possibility of the equilibrium.

 \par
To visualize physics relevant for the shadowing effects we
investigate also the Fourier transform of the ratio of the
distribution function and the invariant energy $s$.
(Within the region of validity of QCD factorization theorem this
ratio has a meaning of
the cross section for the scattering of a dipole off
a target \cite{BBFS}.) The increase with the distance of a Fourier
transform of this quantity shows the existence
of the long range pQCD interaction between the two colorless dipoles.
(Remember that the  long range interaction related to the zero
mass of the gluon is cancelled out in the amplitudes of the
collisions of color neutral objects as the consequence of the
gauge invariance).
\par Let us note that
similar increase with distance can be derived from the calculations of
structure function evaluated within  the Regge pole approximation
\cite{Frishman1,Frishman2} if intercept of Regge pole is
$\alpha (0)> 1$ (Only $\alpha(0) \le 1$ case was considered in
refs. \cite{Frishman1,Frishman2}).

\par Let us recall conventional definitions concerning the
relationship between the products of the currents and the
structure functions. The structure functions are defined through
the current product as
\beq
\frac{1}{\pi}W_{\mu\lambda}(q,p)=-(\delta_{\mu\lambda}-q_\mu
q_{\lambda}/q^2) W_1(x,q^2)+(1/m^2)(p_\mu+q_\mu/(2x))
(p_\lambda+q_\lambda/(2x))W_2 \label{1a} \eeq
\beq
W_{\mu\lambda}=<p\vert\int d^4y \exp(iqy)J_\mu(y)J_\lambda(0)\vert p>
\label{2a} \eeq
Here $J_{\mu}$ is the operator of the
electromagnetic current. These structure functions are usually
redefined into the dimensionless ones:
\beq F_1\equiv W_1,\,\,\,\,\, F_2\equiv (Q^2/2m^2x)W_2 \label{2b} \eeq
Within the DGLAP approximation these structure functions can be
approximated at small $x$ as
\beq
F_2(x,Q^2)=\int^1_{x}dsG_2(x/s,Q^2)g_{\rm gluon}(s,Q_0^2)
\label{3a} \eeq and \beq
xF_1(x,Q^2)=\int^1_{x}(ds/s)G_1(x/s,Q^2)g_{\rm
gluon}(s,Q_o^2) \label{4a} \eeq
The function $g_{\rm gluon}(s,Q_0^2)$ is the nonperturbative gluon
distribution that
parametrizes long distance contributions while the functions $G_i$
describe the  distribution of gluons (sea quarks and antiquarks)
within the gluon. For the gluon distribution similar convolution
formulae are valid. (See e.g. ref. \cite{Sterman} for more
detailed definitions.) Within the parton model the quark -gluon
distribution functions are \beq G_1=G_2=\delta (x-s)
\label{5a} \eeq
\par For the analysis of the light-cone behaviour it is
convenient also to use functions
\beq
V_2=W_2/(m^2_NQ^2),\,\,\,\,\, V_L=((Q^2/x^2)W_2-W_1)/Q^2
\label{i9} \eeq
These functions are free from the kinematic singularities.
\par For the theoretical description of high energy processes
in the nucleon rest frame it is useful to analyze the cross
section \beq \sigma=F_2/Q^2 \label{k} \eeq
instead of the IMF parton distribution- $F_2$:
\par For the gluon-gluon distribution function G this
cross-section has the sense of the dipole-target cross-section
\cite{BBFS}:
\beq
\sigma_d=4\pi\alpha_sx_BG/Q^2\label{cross}
\eeq
\par In the framework of the Feynman parton model
\cite{Ioffe,RE,IKL,Frishman1,Frishman2}
\begin{eqnarray}
V_L(x^2,px)&=&-2\pi
i\epsilon(x_0)\delta(x^2)f_L(px)\nonumber\\[10pt]
V_2(x^2,px)&=&2\pi i\epsilon(x_0)\theta(x^2)f_2(px)\nonumber\\[10pt]
\label{i10}
\end{eqnarray}
while the calculation
based on the Regge models \cite{Regge} gives
\beq
f_L\sim (px)^{\alpha (0)}+\mbox{const},f_2\sim
(px)^{\alpha(0)-2} \label{i11}
\eeq
\par Our main result is the
current-current correlator and
the cross-section in the coordinate space at fixed and sufficiently
small space-time interval $y^2$ but large relative distances $py$ evaluated
in QCD using both leading log and resummation models.
\par The paper is organized in the following way.
In the second chapter we review the results of the parton model
for the structure functions in coordinate space and show in detail how
to account for properly the space-time structure of the commutators
including causality. In the third chapter
we evaluate the light- cone correlators of the currents within the
DGLAP approximation,  and find that at fixed space-time interval they
increase with the distance near the light cone.
We found difficult to calculate Fourier transform of amplitude
directly and to keep causality because of necessity to make
approximations. Instead we generalized method of calculations
developed within parton model in \cite{Ioffe,IKL} .
For this aim we found convenient to use method of moments including
analytic continuation in the vicinity of $n\to 1$.
In the fourth chapter we
evaluate Fourier transform into coordinate space of the
phenomenological and theoretical  gluon distributions in the small
$x$ limit,including both the experimental data and the recent
resummation models. We also  consider the space-time behaviour of
the structure functions if the   Unitarity Bound is achieved at
high energies. In general we find that the rise of the
distributions in the limit $x\rightarrow 0$ leads to the
corresponding rise of the light-cone product of the local
currents. The fifth chapter is the conclusion.

\section{Parton model in the coordinate space.}

\par Let us briefly review the calculations
of the structure functions in coordinate space for $y^2\to 0$
within the parton model \cite{Ioffe,IKL,Frishman1,RE}. Within the
parton model approximation the structure functions are functions
of only $x$. The calculations were carried through in early
seventies assuming dependence on $x$ as given by Regge formulae,
with $\alpha_P(0)\le $. We need to calculate \beq F(y^2,py)=\int
\frac{d^4q}{(2\pi)^3} F(q^2,pq)\exp(iqy) \label{d1} \eeq The
structure function in the parton model can be derived in the
nonperturbative QCD as the discontinuity on the cut in the complex
$pq$ plane: \cite{IKL,Feynman}: \beq
F(q^2,pq)=\int^1_0dx(2pq)\epsilon(pq)(\delta(q^2+2(pq)x)+
\delta(q^2-2(pq)x)F(x)\label{d2} \eeq Here $F(x)$ is the
nonperturbative parton distribution in the target. For simplicity
we consider here spinless quarks. Generalization to spin of quark
$1/2$ is trivial and does not introduce new theoretical phenomena.
Let us briefly review the standard way of the calculation of these
Fourie transforms \cite{IKL}. We shall start from the integral
that is the particular case of the  integral (\ref{d1})-the
integral \beq R(y^2,py)=\int
\frac{d^4q}{(2\pi)^3}\int^1_0dx\epsilon(pq)(\delta(q^2+2(pq)x)+
\delta(q^2-2pqx)\label{d2d} \eeq
 We first calculate the integral
\beq I(y^2,py)=\int
\frac{d^4q}{(2\pi)^3}\epsilon(pq)(\delta(q^2+2(pq)x)+
\delta(q^2-2pqx)\label{d2bu} \eeq
 This integral is equal to a sum of two integrals, each of them corresponds
to the contribution of the region $pq>0$, and $pq<0$: \beq
I(y^2,py)=I^+(y^2,py)+I^-(y^2,py).\label{d2w}\eeq Here
 \beq
I^{\pm}(x^2,px)=\int
d^4q\delta(q^2+2pqx)+\delta(q^2-2pqx)\exp(iqy)\theta(\pm qp)
\label{d3} \eeq
 Making substitution $q\to q+px$ in the first term
and $q\to q-px$  in the second we obtain: \beq
I^{\pm}=2\cos((py)x)\int d^4q/(2\pi)^3 \delta(q^2)\theta(\pm pq)
\exp(iqy ) \label{d4} \eeq The latter integral is well known (see
e.g. ref. \cite{BS}) and immediately gives \beq I^{\pm}(y^2,py)
=2i\cos(x(py))D^{\pm}(y^2,py)). \label{d5s} \eeq Taking the sum we
obtain
 \beq I(y^2,py) =2i\cos(x(py))D(y^2,py)). \label{d5st} \eeq
Integrating now over x we obtain the integral (\ref{d2d})
 \beq
R(y^2,py)=2i \int^1_0 dx  \cos(x(py)) D(y^2,py)\label{d7}\eeq Here
the function \beq D(y^2,py)=
(1/(2\pi))\epsilon(py)(\delta(y^2))-(mx/(2\sqrt{x^2}))\theta(x^2)J_1(mx\sqrt{x^2})
\label{d8} \eeq

is the Pauli-Jordan commutator of the scalar particles (see e.g.
ref. \cite{BS} for detailed analysis of the singular functions in
QFT). We retained the full dependence on the nucleon mass in order
to be sure that there are now singularities in the limit
$m\rightarrow 0$. To account spin of quarks one should substitute
function D by Green function of spin $1/2$ particle -$S(y)$. In
the LT approximation it is necessary to neglect by masses of
quarks. Taking now the $m\rightarrow 0$ limit in the latter
equation we obtain
 \beq
R(y^2,py)=i/\pi \epsilon(py)\delta(y^2)\sin(py)/(py) dy.
\label{d9} \eeq
 We can go now to other structure functions
discussed in the introduction. The corresponding Fourie transforms
differs from the integral (\ref{d2d}) that we had taken by the
powers $(q^2)^n(pq)^m$ where n and m are integer (but generally
non positive numbers). If both n and m are positive we can take
the relevant integrals just by using the corresponding
differential operators. In the parton model the scaling leads to
the general form of the structure function F(x) and one
immediately obtains, acting on the eq. (\ref{d8}) with the
operator $pq=-ip\partial_x$
  that near the light cone
\cite{IKL}
\begin{eqnarray}
F(y^2,py)&\rightarrow&\epsilon(py)/(2\pi)
\int^1_0dx(4i(py)\delta'(y^2)\cos(x(py))-2m^2x\sin(py)\delta(y^2)\nonumber\\
[10pt]&+&
2(p\partial_x)(\theta(y^2)mx/\sqrt{y^2})J_1(xm\sqrt{y^2}))F(x)\nonumber\\[10pt]
\label{d10}\end{eqnarray} Here we once again retain nonzero m to
be sure there are no singularities.
 Taking the limit
$m\rightarrow 0$ we immediately obtain that for the parton model:

\beq F(y^2,py)=(1/\pi)\int^1_0 dx\epsilon(py)
\cos(x(py))\delta'(y^2)(2(py))F(x) \label{d11} \eeq In particular
for $F(x)=1$ we obtain the Fourie transform \beq K(y^2,py)=(2/\pi)
\sin((py))\delta'(y^2)\epsilon(py) \label{d11ab} \eeq
 The integral (\ref{d11}) is well defined for $F(x)\sim x^\alpha$ if
$\alpha
>-1$. This is  however not the general case.
The most interesting structure functions are
$F_1(x)=1/x,V_2(q^2,pq)\sim 1/(q^2 pq)\sim x/q^4, V_L\sim
pq/q^4\sim 1/(xq^2)$. It is easy to see that for these structure
function the integral (\ref{d11}) formally diverges
logarithmically and must be regularised. In order to define these
integrals and satisfy requirement of causality we follow ref.
\cite{IKL}, namely use the differential equations: if two
functions A and B are connected as
$$A=B/q^2$$, then $B(q^2,pq)=q^2A(q^2,pq)$, and in coordinate space we obtain
$$\Box A(y^2,py)=B(y^2,py)$$

Let us use this method for the calculation of the structure
functions defined above. Let us start from $F_1=1/x=-2(pq)/q^2$.
Then one has \beq \Box
F_1(x^2,px)=+2L(pq)=-2i(p\partial_x)K(x^2,px) \label{d13} \eeq
Here L means a Fourie image of the corresponding structure
function. Since $K=4\sin(py)\delta'(y^2)\epsilon(py)$, one obtains
the equation \beq \Box F_1(y2,py= -4i/\pi(py)\epsilon(py)
\sin(py)\delta{''}(y^2)\label{d14} \eeq
We look for the  solution in the form

$$F_1(y^2,py)=A(v)B(u)$$ where $v=y^2,u=py$.
Since $$\Box H(v,u)=4(2F_v+x^2F_{vv}+uF_{uv})$$
where we used $p^2=0$, we obtain
$$\Box (A(v)B(u))=4(2A_vB(u)+vA_{vv}B(u)+uB_uA_v))$$
Note now that \beq v\delta^{n}(v)=-n\delta^{n-1}(v)
\label{d15}\eeq
 as easily
proved by direct calculation. Then if we take $A(v)=\delta'(v)$
one obtains $$\Box H=4(-\delta{''}(v)B(u)+uB_u(u)\delta{''}(v)$$
Comparing this expression with the r.h.s. of eq. (\ref{d14}) we
obtain \beq -i/\pi u\sin(u)=uB_u-B(u) \label{d16} \eeq This
differential equation can be easily solved with the boundary
condition $B(0)\rightarrow 0$. The solution is
\beq B(u)=-i/\pi u\int^u_0ds \sin(s)/s \label{d17} \eeq
Then one obtains \beq F_1\rightarrow
-i/\pi\epsilon(py)\delta'(y^2) (py){\rm Si}(py)\label{d17a} \eeq
Here ${\rm Si}$ is the integral sinus function \beq {\rm
Si}(u)=\int^u_0\sin(s)/sds \label{sinus} \eeq
 (see e.g. ref. \cite{AS} for the detailed review of its
properties). In the limit of large $py$ that we are interested  in
this paper we obtain \beq F_1(y^2,py)\rightarrow
-i/2\epsilon(py)(py)\delta'(y^2)\label{d18}\eeq
Note that the results do not contain any logs as it will follow
naively from the corresponding diverging integral (\ref{d11})
and are causal.

\par Exactly in the same way one can calculate $V_2$ and $V_L$.
For $V_2$ one gets

\beq \Box
V_2(y^2,py)=-i/(2\pi)\epsilon(px)\delta(x^2)\sin(py)/(py)\label{d19}\eeq
This equation can be easily solved using the ansatz

\beq
V_2=\theta (y^2)\epsilon(py)B(py)\label{d20}
\eeq
Repeating the
steps that lead to the solution of the previous equation we obtain
\beq uB_u+B=\sin(u)/u\label{d21}\eeq This equation has the
solution
$$B(u)=\frac{1}{u}Si(u)$$
We immediately obtain
\beq
V_2(y^2,py)= -(i)/(2\pi)(1/py)Si(py)
\theta(y^2)\epsilon(py)\label{d21a}
\eeq
Asymptotically one obtains
\beq
V_2\sim (-i/4)/(py).
\label{d22} \eeq

This is just the result of Ioffe \cite{IKL} (obtained

practically
by the same method). Note that one does not get any large
logs using such method as one will obtain
making naively Fourier transform (see next section). Finally,
using the same approach one can calculate the function $V_L\sim
(pq)/q^4$.
\beq \Box^2 V_L=2(py)\delta^{''}(y^2)\sin(py)\label{d23} \eeq
The general solution is \beq
V_L=(-i/(2\pi))\delta(y^2)\epsilon(py)\cos(py)\label{d24} \eeq
as in ref. \cite{IKL}.

\par Summarizing, in this chapter we reviewed the method
of differential equations due to ref. \cite{IKL} of obtaining the
Fourie transform of the scaling functions, and stressed that this
method permits the one to calculate Fourie transforms without
violating causality. We
 considered the Fourie
transforms of  $F_1$ in the parton model under the condition that
$F_1\sim 1/x$ near
$x\rightarrow 0$,
i.e. has Pomeranchuk behaviour with $\alpha_P(0)=1$.

 \section{Coordinate space representation of the sea quark, gluon
   distribution functions of the gluon.}

\subsection{Fourie transform of the current-current correlators.}

\par In the previous section we
performed the
Fourie transform of
the structure functions in the parton model. Let us now go to the
leading log QCD. Let us make the actual calculation for the
simplest case- Fourier transform into coordinate space of the
structure function of the gluons within the gluon. In the case of
of quark structure function of a quark or a gluon all calculations
are practically identical. So there is no need to repeat them. All
calculations will be made in the target rest frame because the
space-time evolution is most straightforward in this frame. Our
calculations will be legitimate in the limit of the fixed

space-time interval $y^2$ but $py\to \infty$. We choose this limit

because there exists rather direct correspondence between the
structure functions at small x and Fourier transform. We start
from the expression for the gluon structure function which is the
solution of the DGLAP equation in the double logarithmic
approximation \cite{DGLAP,Doc}.  To derive analytic formulae  we
neglect the running of the coupling constant which should be slow
because of the smalness of
$y^2$
The DGLAP equation is \beq
Q^2\frac{d}{dQ^2}G(z,Q^2)=(\alpha_s/(2\pi ))\int^1_z(dz'/z')
\gamma_{GG}(z/z')G(z',Q^2). \label{1} \eeq Here $\gamma_{GG}$ is
the kernel in the QCD evolution equation, \beq \gamma_{\rm
GG}=2N_c/z\label{ker}\eeq
 The solution of
this equation is given by
\beq
G(z,Q^2)=\int dn/(2\pi i)(z_0)^{n-1}/z^n(Q^2/Q_0^2)^{\gs} \label{2}
\eeq
Here the contour
integration over n runs along a straight line parallel to the
imaginary axis to the right of all singularities of the integral.
We use the notation $Q^2=-q^2$ if $q^2\le 0$ and $Q^2=q^2$ if
$q^2\ge 0$. This expression can be rewritten using the scaling
variable \beq x_B=-q^2/(2pq) \label{3} \eeq The above solution
corresponds to simple initial conditions
$$x_BG(x_{B},Q^2)=\delta (x_{B}-x_{0B})$$
\par We shall need the Fourie transform of the eq. (\ref{2}).
\beq G(py,y^2)=\int d^4qG(x,Q^2)\exp(iqy) \label{4} \eeq \par
Let us first determine the integration area. The structure
function is
symmetric between the u and s channels. So ${\rm Im}_s F_1={\rm
Im} F_u$.
This means the invariance on the substitution 
$x\rightarrow -x$,
$pq\rightarrow -pq$ . Thus
one can limit the area of integration by the region
($pq\ge 0$).
In this area there are four subregions:
\par i. $q^2\ge 0$ and
$0\ge x_B\ge -1$
which corresponds to the $e^+e^-$ fragmentation into hadrons in the
field of target. Within the LL approximation this structure function
is zero.

ii. $q^2\ge 0$ and
$x\le -1$
Amplitude in this kinematics can be related with the inclusive
process :$e^+e^{-}\to N+X$. We will show that this region gives no
significant contribution to the kinematics of interest in this
paper. In the LL approximation this amplitude is connected to DIS
amplitude by Gribov$-$Lipatov relation (see below).

iii.$q^2\le 0$, and $1\ge x\ge 0$
that corresponds to the deep-inelastic scattering (DIS).

iv.$q^2\le 0$, and
$x\ge 1$.
In this area the structure function
is 0 because of the energy-momentum conservation laws.
\par In the
second region one has an additional kinematical restriction
\beq
x \ge 1, q^2\ge 4m^2x^2 \label{5} \eeq
that just expresses the condition $q_O^2\ge \vec q^2$ in this
kinematical area.
\par In the  third region it is worthwhile to use instead of $q^2$
as an invariant variable $Q^2=-q^2$, and it is easy to see that
kinematically
$x\le 1$.
\par We shall start from the DIS region.
\par
Naively, in order to carry the Fourie transform in eq. (\ref{4}),
one can use the GIP approximation \cite{GIP,IKL}. In this
approximation one takes into account that the integrand in the
laboratory reference frame is dominated by \beq q_0^2\ge
(q^2)^2/(4m^2)\gg \vert q^2\vert \label{t1}\eeq Correspondingly,
one can expand $\sqrt{q_0^2-q^2}\sim q_0-q^2/(2q_0)$. Using this
approximation one obtains the integrand directly as a function of
$y^2$.
The arising integrals can be easily calculated. However,
they do not satisfy the causality condition: the commutator is
nonzero for
$y^2\le 0$,
and this condition must be imposed by
hand. It is easy to see, taking as an
pattern
the calculations from the previous chapter for the parton model
and trying to do them explicitly calculating the integrals in the
GIP approximation, that the problem is the limitation of the
integration area by the condition (\ref{t1}). Then even when we
obtain the convergent integrals it is not clear how to obtain
casuality naturally. Instead, we shall adopt here a different
approach. It is possible to prove that the Fourie transform of
$(q^2)^n/x^m$
is the analytical function of $n$ , uniquely defined by its values
in integer n, where the latter function is understood as a
generalized one. Let us start from the integral (\ref{d11ab}) and
multiply the integrand by $(q^2)^n$. The integral is obtained by
acting with the operator $\Box^n$ on the result of the
integration. For the leading term in the asymptotics
in $py$
one obtains:
\beq (q^2)^n\rightarrow \Box^n(2/\pi)(-1)^n \int^1_0dx
\cos(x(py))\delta'(y^2)(py)\label{t3}\eeq
After differentiating one gets
\beq q^{2n}F(x)\rightarrow \int^1_0dx
2^{2n}(py)^{n+1}\delta^{n+1}(y^2) x^n\cos(x(py)+n\pi/2)F(x)
\label{t4} \eeq
There exists however the unique generalized function such that it
is an analytic function of n and is equal to $\delta^{n}(u)$ for
positive integer n \cite{Gelfand}. This function is \beq
J(s,u)=u_+^{s-1}/\Gamma(s)\label{t5}\eeq For this function \beq
\lim_{s\rightarrow -n}J(s,u)=\delta^{n}(u)\label{t6}\eeq and we
denote $u_+\equiv \theta(u)u$ (the standard notation in the
mathematical literature \cite{Gelfand}.)
 Then we can extend eq. (\ref{t4}) for
noninteger n as
 \begin{eqnarray}
q^{2s}F(x)/2^{-2s}&\rightarrow&\int^1_0dxx^{-s}(py)^{-s+1}
\cos(x(px)-s\pi/2))F(x)\nonumber\\[10pt]
&\times&(x^2_+)^{-s-2}/\Gamma(-s-1)\nonumber
\\[10pt]\label{t6a}\end{eqnarray} Here
$x^2_+=\theta(x^2)x^2$.
\par Let us shortly discuss the result from the mathematical point
of view. It is easy to check by using the inverse operator of
Laplace as we did in the previous section that the asymptotics
(\ref{t6a}) is valid for negative n. Thus we have the problem of
restoring function that is known for all integer n, analytical in
n and has power like asymptotics. Such function is uniquely
defined \cite{FKP}, as it is well known from the theory of complex
variables (and Regge calculus, where the corresponding procedure
is called Gribov-
Froissart
projection). Then eq. (\ref{t5}) fully defines the function. It is
straigtforward to see  that this function coincides with the one
obtained in GIP approximation, except an important difference: we
automatically achieve causality. Thus our approach-analytically
continuing the result of the parton model is the only possible
approach to the Fourie transform. Once we know how to deal with
the powers of $q^2$, we must put $s=\gs$ and carry the remaining
integration over x. We obtain using eq. (\ref{t6}):
\begin{eqnarray}
G(px,x^2)&=&(\pi/(2py))\int dn/(2\pi i)\int^1_0dx/x^2
\cos(-(\gs+2)\pi/2+xpy))\nonumber\\[10pt]
&\times&\Gamma(\gs+2)(x)^{\gs+2-n}/\vert
(x^2/(2px))\vert^{\gs+2}\nonumber\\[10pt]
&\times&(Q_0^2)^{-\gs}z_0^{n-1}(2m)^{\gs+2}\nonumber\\[10pt]
\label{17}
\end{eqnarray}
The
integration in the above formulae can be easily performed.
\cite{AS,BE,GR}:
\begin{eqnarray}
\int^1_0\sin(a+mxr)x^{\mu-1}dx&=&(\sin(a)(F(imr)+F(-imr))\nonumber\\[10pt]
&-&i\cos(a) (F(imr)-F(-imr)))/(2\mu).\nonumber\\[10pt]
 \label{18} \end{eqnarray} Here the
function F is the confluent hypergeometric function: \beq
F(x)=_1F_1(\mu,\mu+1,x),\,\,\,\,\,F(0)=1 \label{19} \eeq We are
actually interested in the limit of large distances (times),
$r\rightarrow\infty$. In this limit one uses the asymptotics:

\beq F(imr)=\Gamma
(\mu+1)\exp(i\pi\mu)/(imr)^\mu+\mu\exp(imr)/(imr) \label{20} \eeq
\beq F(-imr)=\Gamma
(\mu+1)\exp(-i\pi\mu)/(-imr)^\mu+\mu\exp(-imr)/(-imr) \label{20a}
\eeq or, since \beq F(imr)+F(-imr)=2(\Gamma(\mu+1)\cos(\pi
\mu/2)/(mr)^\mu+\mu\sin(mr)/(mr)) \label{p1} \eeq and \beq
F(imr)-F(-imr)=2i(\Gamma(\mu+1)\sin(\pi
\mu/2)/(mr)^\mu-2\mu\cos(mr)/(mr)) \label{p2} \eeq we have \beq
\int^1_0dxx^{\mu-1}\sin(a+mxr)\sim
2(\Gamma(\mu)\sin(a+\pi\mu/2)/(mr)^\mu- \cos(mr+a)). \label{p3}
\eeq
 Using the latter integral we finally obtain the contribution
from the deep-inelastic scattering region in the Fourie transform.
In our case \beq a+\mu\pi/2=2\gs\pi/2-(n+1)\pi/2\label{21a}\eeq
\beq
 \mu=\gs-n+1,a=-(+\gs+1)\pi/2 \label{21} \eeq Then we have
\begin{eqnarray}
G(py,y^2)&=&(\pi/(py))\int dn/(2\pi i)\Gamma(\gs+2)\nonumber\\[10pt]
&\times&(+\cos(py-(\gs)\pi/2)/(mr)\nonumber\\[10pt]
&+&
\Gamma(\gs-n+1)\sin(n\pi/2)/\nonumber\\[10pt]
&/&((py)^{\gs-n+1}\vert (y^2/(2py))\vert^{\gs+2})(Q_0^2)^{-\gs}\nonumber\\[10pt]
&\times&z_0^{n-1}\nonumber\\[10pt]
\label{22}
\end{eqnarray}

\par We have found the contribution due to the DIS into the integral over n.
The contribution due to the fragmentation of $e+e^{-}$ in the
nucleon color field into the integral over n is zero in the
leading log approximation (see above and ref. \cite{GL}).
\par Let
us now consider the contribution due to the annihilation. The
current commutator in the annihilation region can be expressed
through the correlation function in DIS via the Gribov-Lipatov
relation \cite{GL}
\beq G^a(z,q^2)=zG(1/z,Q^2) \label{23}
\eeq
Here $z$ is the Bjorken variable defined in the same way as
$x$, only for the different kinematic region.
\begin{eqnarray}
G(py,y^2)&=&(\pi/(2py))\int dn/(2\pi i)\int^\infty_1z
\cos(+(\gs+2)+py z))\nonumber\\[10pt]
&\times&\Gamma(\gs+2)(z)^{\gs-n}/\vert
(y^2/2py))\vert^{\gs+2}\nonumber\\[10pt]
&\times&(Q_0^2)^{-\gs}z_0^{n-1}\nonumber\\[10pt]
\label{25}
\end{eqnarray}
The integral over $y$ can be easily taken using the integral
\beq \int^\infty_1\sin(a+mrz)z_b^{\mu-1}dz=\int^\infty_0
\sin(a+mrz)y^{\mu-1}dz-\int^1_0\sin(a+mrz)z_b^{\mu-1}dz
\label{26} \eeq Then first integral is taken explicitly, while the
second was taken above: \beq
\int^\infty_1\sin(a+mrz)z^{\mu-1}dz
=\Gamma(\mu)\sin(\mu\pi/2+a)-\int^1_0\sin(a+mrz)z_b^{\mu-1}dz
\label{27} \eeq The asymptotic expansion of the latter integral is
known, and it is straightforward to see that for large $r$
\beq\int^\infty_1\sin(a+mrz)z^{\mu-1}dz\sim +\cos(mr+a)/mr
\label{28} \eeq The reason that the latter integral does not
depend on $mu$ is that the asymptotics is dominated by the area
$z\sim 1$, where
$$z^{\mu-1}=\exp((\mu -1)\log(z))\sim 1.$$
It is  clear that the similar term in eq. (\ref{17}) also comes
from the area $x\sim 1$, and these terms correspond to the
contribution of the parton model. The two contributions are very
similar with the only difference that we must use in eq.
(\ref{28}) \beq a=-(\gs+2)\pi/2 \label{29} \eeq Then we can write
the expression for the structure function that includes both the
DIS and annihilation regions:
\begin{eqnarray}
G(py,y^2)&=&(\pi/(mr))\int dn/(2i\pi) \Gamma(\gs+2)
)\nonumber\\[10pt]
&(&+(\sin(py)\cos((\gs)\pi/2)\nonumber\\[10pt]
&+&
\Gamma(\gs+2)\Gamma(\gs-n+1)\nonumber\\[10pt]
&\times&\sin(n\pi/2)/(py)^{\gs-n+1}\nonumber\\[10pt]
&/&\vert (y^2/(2py))\vert^{\gs+2}(Q_0^2)^{-\gs}\nonumber\\[10pt]
&\times&z_0^{n-1}\nonumber\\[10pt]
\label{30}
\end{eqnarray}
\par We see that the current-current correlator contains two
distinct contributions. The first is due to
$x\sim 1$ .
The second is solely due to perturbative gluon effects. This part is
dominated by moderately small
$x$.
\par We can now take an integral over n. Let us start from the
$x\sim 1$ contribution
\begin{eqnarray}
G_1(py,y^2)&=&-(1/(2mr)))\int dn/(2\pi i)\cos((\gs)\pi/2)
\sin(mr)/(mr)\nonumber\\[10pt]
&\times&\Gamma(\gs+2)/\vert
t-r\vert^{\gs+2}(Q_0^2)^{-\gs}\nonumber\\[10pt]
&\times&z_0^{n-1}(2m)^{\gs+2}\nonumber\\[10pt]
\label{30a}
\end{eqnarray}
The second contribution is due to the moderately small
$x$
It is equal to
\begin{eqnarray}
G_2&=& \Gamma(\gs -n+1)\Gamma(\gs+2)\nonumber\\[10pt]
&\times&\sin((n)\pi/2)\nonumber\\[10pt]
&/&((mr)^{\gs-n+1}\vert t-r\vert^{\gs+2})(Q_0^2)^{-\gs}\nonumber\\[10pt]
&\times&z_0^{n-1}(2m)^{\gs+2}\nonumber\\[10pt]
\label{31}
\end{eqnarray}
These integrals can be taken using the saddle point method.
Consider first the
contribution where  $xG\approx const$ for small $x$
to honor distinctive property of  soft QCD
amplitudes to  significantly more slowly increase with energy as
compared to the amplitudes of hard processes. This contribution
can be studied using the saddle point approximation. Indeed, we have
\begin{eqnarray}
G_1(py,y^2)&=&\int dn/(2\pi
i)(1/y^2)(\cos(py)\cos(\gs\pi/2)\nonumber\\[10pt]
\nonumber\\[10pt]
(&2&mr/(y^2Q_0^2))\cos(\gs\pi/2)\Gamma(\gs+2)z_0)^{n-1}\nonumber\\[10pt]
\label{31a}
\end{eqnarray}
The saddle point is at \beq n-1=\sqrt{(\alpha_sN/\pi)
\log(2py/y^2Q_0^2)/\log(z_0)} \label{31b} \eeq
We then obtain:
\begin{eqnarray}
G_1&=&(\cos(py)/(y^2)^2)\cos(\sqrt{\alpha_sN_c/\pi)\log(z_0)/\log(2py/y^2Q_0^2)}
\nonumber\\[10pt]
&\times&\Gamma(2+\sqrt{\alpha_sN_c/\pi)\log(z_0)/\log(2py/y^2Q_0^2)})\nonumber\\[10pt]
&\times&
\exp(\sqrt{\alpha_sN_c/\pi\log(mr)\log(z_0)\log(y^2Q_0^2)})\nonumber\\[10pt]
&\times&(\alpha_sN_c/\pi)^{1/4}\log(2py/y^2Q_0^2)^{1/4}/\log(z_0)^{3/4}
\nonumber\\[10pt]
\label{31c}
\end{eqnarray}
The last line corresponds to the preexponential.
\par Consider now the second integral (\ref{31}).
The integral can be rewritten in dimensionless variables
as\begin{eqnarray} G_2&=&
\Gamma(\gs -n+1)\Gamma(\gs+2)\sin((n+1)\pi/2)(py)^{n+1}\nonumber\\[10pt]
&/&(Q_0^2x^2)^{\gs+2}\nonumber\\[10pt]
&\times&z_0^{n-1}(2)^{2(\gs+2)}\nonumber\\[10pt]
\label{31d}
\end{eqnarray}
 We immediately see that the saddle point is determined from the
equation \beq \log(pyz_0)=-\alpha_sN_c/(\pi (n-1)^2)\log(Q_0^2x^2)
 \label{32} \eeq We obtain \beq
n-1=\sqrt{(\alpha_sN_c/\pi)(\log(1/Q_0^2x^2))/\log(mrz_0))}
\label{33} \eeq Then substituting in the integral one immediately
obtains the asymptotics
\begin{eqnarray} G_2&=&
\Gamma(\sqrt{(\alpha_sN_c/\pi}(\sqrt{\log(pyz_0)/\log(y^2Q_0^2)}-
\sqrt{\log(mrz_0)/\log(y^2Q_0^2)})\nonumber\\[10pt]
&\Gamma&(\sqrt{(\alpha_sN_c/\pi)( \log(mr)/\log(y^2Q^2_0)}+2)
)\sin(\sqrt{\alpha_sN_c\pi\log(Q^2_0y^2)/\log(pyz_0)/2))}\nonumber\\[10pt]
&\times&(\alpha_sN_c/\pi)^{1/4}
\displaystyle{\frac{\log(Q_0^2y^2)^{1/4}}{\log(py)^{3/4}}}\nonumber\\[10pt]
&\times&(1/y^2)^2\exp(2\sqrt{\alpha_2N_c/\pi
\log(mr)\log(Q_0^2x^2))})\nonumber\\[10pt] \label{w}
\end{eqnarray}
\par Let us now check the applicability of the saddle point method.
It is easy to see that the condition is \beq
\alpha_s\log(py)\log(y^2Q_0^2)\gg 1 \label{34} \eeq Thus the
equations (\ref{32}),(\ref{33}), (\ref{w}) are not valid in the
limit $\alpha_s\rightarrow 0$ that corresponds to parton model.
\par We have found the asymptotics of the current-current
correlator in the double-logarithmic limit of QCD. In this limit
the saddle point method is applicable and the correlator increases
with distances. The applicability condition of this method is
evidently the existence of two large logarithms: the parameter
$$\alpha_sN_c/\pi\log(Q_0^2y^2)\log(py)\gg 1$$.
\par Note, that, generally speaking it is beyond of the accuracy
of the method to keep single logs in the arguments of the
exponents in the above expressilons, and the
legitimate
answer for asymptotics
\beq G^{\rm DGLAP}(py,y^2)=\theta(y^2)\frac{py}{(y^2)^2}
(\alpha_sN_c/\pi)^{1/4}
\displaystyle{\frac{\log(Q_0^2y^2)^{1/4}}{\log(py)^{3/4}}}
\exp(2\sqrt{\alpha_sN_c/\pi \log(py)\log(Q_0^2y^2))})  \label{w1}
\eeq Here we put all terms with single log in the arguments of
exponents to 1.Note that the leading asymptotics is given by the
integral (\ref{w}).
\par Thus we obtained asymptotics for perturbative QCD.
Note that delta-function singularities on the light cone for $F_2$
and $F_1$ were translated into $1/(x_+^2)^2$ behaviour in the
perturbative QCD.
\subsection{Coordinate space physics relevant for cross sections .}
In the previous section we discussed the space-time asymptotics of
the current-current commutator in the LL approximation of PQCD.
This commutator has a well defined probabilistic interpretation in
the infinite momentum frame .  However in the target rest frame
significantly more direct interpretation has cross section
for the dipole scattering of a target and
related shadowing effects . The cross-section is equal to the
correlator divided by an invariant energy,i.e. by s, which means
the commutator must be multiplied by $2x/Q^2$. Thus, in the
notations of the introduction we have to calculate
\beq D(py,y^2)
=\int d^4q\exp(iqy)G(y,Q^2)/(s)\equiv\int
d^4q\exp(iqy)(y/Q^2)G(y,Q^2) \label{41} \eeq Here $s=Q^2/y$ is the
invariant energy squared.
\par This quantity
is sometimes considered as
a "potential" for the interaction between color-neutral dipoles.
For this quantity we may repeat the analysis of the previous
section. It is straightforward to see that both for parton model
and large Bjorken $x$, large $Q^2$ regime of DGLAP equations, the
effect is the loss of one power of
$y^2$, in the denominator and the loss of one power
of (py) in the nominator, i.e. the correlator increases
logarithmically in the parton model (see previous subsection). For
the perturbative QCD asymptotics we obtain:
\beq D(py,y^2)=\theta(y^2)\frac{1}{y^2} (\alpha_sN_c/\pi)^{1/4}
\displaystyle{\frac{\log(Q_0^2y^2)^{1/4}}{\log(py)^{3/4}}}
(1/x^2)^2\exp(2\sqrt{\alpha_sN_c/\pi
\log(py)\log(Q_0^2y^2)})\label{w21a} \eeq
The function D thus
increases in the perturbative QCD. \subsection{More about parton
model.}
\par The modern definition of the parton model ordinary refers
to nonperturbative distributions without taking into account
perturbative QCD evolution, i.e. for initial conditions for
evolution equations, such as DGLAP. The most popular form of the
initial conditions (that also gives a best agreement with the
experimental data) is \beq F_2(x)=C/x^{\alpha} \label{z1} \eeq
where $\alpha >0$. This case was not considered in early seventies
since that time it was assumed that $\alpha\le 0$ for the physical
cases.
 The Fourie transform of this function, except
the special case of the integer $\alpha$  is clearly given by the
analytic continuation of the asymptotics obtained in the previous
section. If we continue analytically the equations from the last
section and use the results of the section 2 we immediately obtain
\begin{eqnarray} F_2&\rightarrow& \int^1_0dx\cos(x(py))\delta'(y^2)(py)
x^{(1-\alpha)-1}\nonumber\\[10pt]
&\sim&
2(\Gamma(1-\alpha)\sin(\pi\alpha)(py)^{\alpha}\nonumber\\[10pt]
\label{z2}
\end{eqnarray}

Note that for $\alpha=0$ this term becomes zero and the
asymptotics will be given by oscillating term 
$\sim \sin(py)/(py)$
(times the same type of light-cone singularity).
\par For the structure function $F_1$ the behaviour is $\sim
x_B^{-(\alpha +1}$, and one continues analytically eq. (\ref{z2})
obtaining the increase of the commutator as
\beq F_1\rightarrow 2(
\Gamma(-\alpha)\cos(\pi\alpha/2)(py)^{\alpha+1}\delta'(y^2)
\label{z3} \eeq
This expression has a pole singularity for
$\alpha=0$, when we return to the function already considered in
the framework of the parton model.
\section{Phenomenological distributions}
\subsection{Small
structure functions.}
\par In the previous section we analyzed the space-time structure
of the correlators due to a parton model and within the area of
applicability of leading order DGLAP equations. Including NLO will
not change our conclusions. However, at extremly small x
(kinematics of LHC?) where energy conservation law does not
preclude large number of gluon radiations in the multiRegge
kinematics
$x$
limit of the DGLAP equations is, literally speaking, not available
and instead one needs to use for G either by resummation
approaches \cite{ABF,Ciafaloni}, or phenomenological ones or
phenomenological one from HERA \cite{Zeus}:
\beq
G_p(x,Q^2)\sim (1/x)^{\alpha+1}(Q^2)^\beta \label{43} \eeq
where \beq \alpha\sim 0.25,\beta\sim 0.25 \label{44} \eeq The
theoretical distributions expected within the resummation
approaches of \cite{ABF,Ciafaloni} differ from a phenomenological
one by logarithmic terms:
\beq G_T(x,Q^2)\sim (1/x)^{1.25}(Q^2)^{0.25}
/\sqrt{\log(x)^3}\label{43a} \eeq It is straightforward to carry
 Fourie transform of this distribution. Once again we carry the
analytical continuation of the parton model formulae. The
singularities $\delta(y^2)$ and $\delta'(y^2)$ are smoothened into
$1/(y_+^2)$ and $1/(y_+^2)^2$ respectively.
 We obtain \begin{eqnarray}
G(py,y^2)&=&\pi\Gamma(\beta
+2)(py)^{\beta+1}/(y^2)^{2+\beta}\int^1_0dx
x^{\beta-\alpha-1}\sin(x(py)-\beta\pi/2)/(py)\nonumber\\[10pt]
D(py,y^2)&=&\pi
\Gamma(\beta
+1)(px)^{\beta}/(x^2)^{1+\beta}\int^1_0dx
x^{\beta-\alpha-1}\cos(x(py)-\beta\pi/2)/(py)\nonumber\\[10pt]
\nonumber\\[10pt]
\label{nb}
\end{eqnarray}
The relevant asymptotics is obtained from the asymptotics of the
confluent hypergeometric function as in the previous section . For
$\beta-\alpha\le 0$ the integrals must be considered in the
analytic continuation sense, for $\beta=\alpha$ one obtains the
logarithmic divergence that must be dealt with as in the parton
model in the previous subsection.
\par  Altogether one obtains, if $\beta\ne\alpha$.
\beq G(py,y^2)=\theta(y^2)\pi\Gamma(\beta+2)\Gamma(\beta-\alpha)
\sin(\pi\alpha/2))((py)^{\alpha+1}/(y^2)^{\beta+2} \label{44a}
\eeq and \beq
D(py,y^2)=\theta(y^2)\pi\Gamma(\beta+2)\Gamma(\beta-\alpha)(1/2)
\cos(\pi(\alpha)/2)((py)^{\alpha}/(y^2)^{\beta+1} \label{45} \eeq
  If $\beta=\alpha\ne 0$ one obtains logarithmic
 asymptotics:
\beq G(py,y^2)=\pi\Gamma(\beta+2)
\sin(\beta\pi/2)((py)^{1+\beta}\log(py)/(y^2)^{\beta+2} \label{46}
\eeq
 and \beq D(py,y^2)=\pi\Gamma(\beta+2)\cos(\beta\pi/2)
(py)^\beta\log(py)/(y^2)^{\beta+1} \label{47} \eeq For the HERA
phenomenological case one has $\beta\sim\alpha$, the same is true
for recent phenomenological asymptotics due to refs.
\cite{ABF,Ciafaloni}. Thus for them we obtain the logarithmic
times power increase of D and G functions on the light cone. \par
It is interesting to note that recently Ciafaloni et al
\cite{Ciafaloni} suggested resummation model where the structure
function may have a dip in the energy dependence, postponing
increase to smaller x than in the kinematics of HERA. This will
postpone increase of D to larger x. Additional factor in the
asymptotics $\sim 1/\log^{3/2}(1/x)$ claimed in
ref\cite{Ciafaloni} may change the behaviour of D, making it
slowly increasing with distance as $(py)^{0.2}/\sqrt{\log(py)}$
and having a dip for some interval of py.
\par In the case of the leading order BFKL approximation
we have $\beta=1/2$, $\alpha\sim 0.8$, and we have asymptotics

\begin{eqnarray}
G(px,x^2)&\sim& (py)^{1.8}/(y^2)^{5/2}\nonumber\\[10pt]
D(py,y^2)&\sim&(py)^{0.8}/(y^2)^{3/2}\nonumber\\[10pt]
\label{50}
\end{eqnarray}
\subsection{Black limit}
\par It is worth to analyze coordinate space dependence for Unitarity Limit
for structure functions for the small x behaviour of
structure functions \cite{MacDermot}  Unitarity Bound=
black body approximation \beq G(x,Q^2)\propto (1/x)(Q^2/Q_0^2) \log^3(x_{0}/x)
+\mbox{ peripheral=DGLAP terms}
 \label{51}
\eeq and \beq \sigma (s)\sim \log^3(x/x_{0}) \label{s1}\eeq
where $x_0$ is weak function of $Q^2$. Doing Fourie transform we
obtain

\beq
 G(py,y^2)\propto (py)\theta (y^2)\log^3((py)))/(y^2)^3 +\mbox{
peripheral terms} \label{52} \eeq \beq D(py,y^2)\propto \theta
(y^2) \log^3((py))/(y^2)^2) +\mbox{ peripheral terms} \label{53}
\eeq

Thus in the black limit both G and D at given impact parameter
contain trivial increase with distance:factor r is because the
same dipole is probed at different space-time points, one
$ln(1/x)$ is due to ultraviolet divergence of renormalization of
e.m. charge. ($\ln^2(x_0/x)$ is due to increase with energy of
impact parameters in the scattering process. Excluding above
factors we find that increase of commutator with distance is
stopped within this limit.
   Structure function continues to increase with energy for the
configuration in the photon wave function  with $4k_t^2\ge Q^2$ as
the consequence of renormalizability of QCD,and because of
increase with energy of essential impact parameters.
\subsection{Hard diffractive processes}
\par Another interesting question is a question of
space-time evolution of hard diffractive processes. These are the
processes $\gamma^*+p\rightarrow X+p$ where X is vector meson,dijet etc.
For this process the relevant amplitude behaves as
\beq
A\sim (s/Q^2)1/Q^n
\label{f1}
\eeq
where n=1/2 or 1. Repeating calculations as above we obtain
that coordinate space amplitudes increase with distances as amplitudes
of LT processes. However dependence on $y^2$ will be weaker by the
factor $(y^2)^{n}$.

\section{Conclusion.}
\par We have studied the
dependence of the current-current correlators (gluon distributions)
in coordinate space
on  $py$ at fixed $y^2$ close to light-cone.
Quite surprisingly, we found that
all theoretical approaches (DGLAP, BFKL, recent resummation models
of small $x$ behaviour \cite{ABF,Ciafaloni}, Unitarity Bound
\cite{MacDermot} and phenomenological description of data, all
lead to the increasing with the distance current-current
correlators, \beq G(py,y^2)\sim py/(y^2)^2)(py)^\alpha/(y^2)^\beta
\label{end}\eeq here indexes $\alpha$ and $\beta$ are  model
-dependent
 at present
but positive. The DGLAP equations
in double logarithmic approximation lead to increase
\beq
 \sim \theta (y^2)(py)/(y^2)^2(\alpha_sN_c/\pi)^{1/4}
\displaystyle{\frac{\log(Q_0^2y^2)^{1/4}}{\log(py)^{3/4}}}
\exp(2\sqrt{\alpha_sN_c/\pi \log(py)\log(Q_0^2y^2)})
 \label{end1}
 \eeq
 Thus increase the current-current
correlators with distance near light-cone is the general feature
of high-energy scattering processes. Moreover, we see that, apart
from the kinematical multiplier $py$ this feature appears due to
the interaction with the gluons and is absent in the parton model,
where only fixed (except the kinematical multiplier $py$
)amplitude oscillations occur.
\par The increase  of the commutator with the distance at the light cone is
relativistic effect, present in the Minkowsky space only, and
it is absent in the Euclidean space.
\par This feature is closely connected
with the known increase of the correlation length at high energies (the
"Ioffe time", \cite{Ioffe,KS}).
\par Such increase seems to be a characteristic feature of a
Pomeron, i.e. of a contribution into amplitudes of a t-exchange
with vacuum quantum numbers. Fourier transform of amplitude with
nonvacuum quantum numbers in t channel (contribution into cross
section) decreases with distance as $(1/r)^{\alpha(0)-1}$. Here
$\alpha$ is intercept of trajectory of dominant Regge pole contribution.
\par Let us stress that we consider the asymptotics near the
light-cone, i.e. $r\sim t$. On the other hand, for equal time
commutator,i.e. small time t but $r\rightarrow \infty$
dominant contribution into Fourier transform arises due
to the region of large $q_0\sim 1/t$ and small space momenta. i.e
the region around x=1 but $q^2\rightarrow \infty$. In this region
Fourier transform oscillates.

\par We can also evaluate product of 4 currents in the same way and to
obtain similar results as above

j(y)j(z)j(0)j(0) in the kinematics

$$y^2\rightarrow 0, z^2\rightarrow 0,  r_y,r_z\rightarrow
\infty.$$
This correlator appears  in heavy ion collisions as a correlation
function between two hard processes which occur at different
space-time points.
\par Finally, we found that the
"potential" function $D(y^2,py)$ that has a physical sense of a dipole
-target "potential"
increases with $py$, in all
above
cases.
This increase
disappears for
the Fourie transform of the Unitarity Bound formula after
excluding effects beyond long-range dynamics. Thus we have found
another instability of the description of the physical state in
terms of quarks and gluons. The physical consequences of this fact
will be discussed in more details in the future publications.
\acknowledgements {The authors thank M. Strikman for very usefull
discussions. }

\end{document}